\documentclass[aps,prb,amsfonts,amssymb,twocolumn,amsmath,preprintnumbers,nofootinbib,floatfix,showpacs]{revtex4}
\usepackage{amsmath,bm,amsfonts,amssymb}
\usepackage[dvips]{graphics}
\usepackage{graphicx}

\begin{document}

\title{Discrete-lattice model approach to the tunneling between d-wave superconductors:
interference of tunnel bonds}

\author{A. M. Bobkov}
\email[Electronic address: ]{bobkov@issp.ac.ru}
\affiliation{Institute of Solid State Physics, Chernogolovka,
Moscow reg., 142432 Russia}

\date{\today}

\begin{abstract}The Josephson current between d-wave superconductors
is investigated in the framework of tight-binding lattice model.
The junction is modelled by a small number of connecting bonds. It
is obtained that the Josephson current through one bond vanishes
when at least one of the superconductors has $(110)$
interface-to-crystal orientation. Interference between the nearest
bonds is appeared to be very important. In particular, it is the
interference term that leads to the nonzero Josephson current for
$(110)$ orientation. Also, interference of two connecting bonds
manifests itself in non-monotonic behavior of the critical
Josephson current in dependence on the distance between the
bonds.
\end{abstract}

%%% PACS numbers

\maketitle

Electronic transport through the junctions between high-$T_c$
cuprate superconductors has been the object of interest for many
years. In particular, the dc Josephson effect has been intensively
studied
theoretically\cite{Barash96,Tanaka96,Tanaka97,Shumeiko01,Shirai03}.
These investigations were inspired by the fact that
superconducting order parameter dominantly exhibits d-wave
symmetry in high-$T_c$ cuprates. This was established by
SQUID-like experiments\cite{Wollman93} and the tricrystal
experiment\cite{Tsuei94}. Theoretical investigations predicted a
number of results, which are the consequences of d-wave nature of
superconducting order parameter. For example, the zero-energy
Andreev bound state (ZES) is formed at junction
interfaces\cite{Hu94}, what gives rise to anomalous enhancement of
the critical Josephson current in d-wave superconductor junctions
at low temperatures\cite{Barash96,Tanaka96,Tanaka97}. Also, for a
mirror junction, where the order parameter on the both sides of
the junction is rotated by the same angle in opposite directions,
a non-monotonic temperature dependence of the critical current was
predicted\cite{Barash96,Tanaka96,Tanaka97}. The temperature
dependence of the critical current was studied experimentally on
the grain boundaries with well-defined lattice
orientation\cite{Dimos90,Hilgenkamp98,Arie00,Il'ichev01}. For
mirror junctions the non-monotonic behavior was found in
\cite{Il'ichev01}, whereas in other cases a monotonic behavior was
reported\cite{Dimos90,Hilgenkamp98,Arie00}. In addition, for the
$45^\circ$ asymmetric junction a $\sin 2\varphi$-like
current-phase relation was predicted\cite{Tanaka97}. However, in
the experiments on the asymmetric junctions not all samples
exhibit the predicted $\sin 2\varphi$-like current-phase
relation\cite{Il'ichev01,Il'ichev98,Il'ichev99}.

The theoretical investigations of the Josephson current for the
junctions between d-wave superconductors were carried out on the
basic of continuous approach\cite{Barash96,Tanaka96,Tanaka97} as
well as making use of tight-binding lattice model\cite{Shirai03}.
The main results of these methods are consistent with each other
for the case of planar junctions. However, the lattice-model
approach can not only give a more realistic description of the
electronic structure of the copper oxide planes of high-$T_c$
superconductors and allows to mimic the corresponding Fermi
surfaces, but it also gives the possibility to study electronic
transport through quantum point contacts of various types. The
recent advances in the fabrication of nanoscale
devices\cite{vanderPost94,Scheer97,Scheer98,Ludoph00} (for a
review see \cite{Agrait03}) has provoked a renewed interest in the
detailed analysis of models involving a few conducting channels.
Josephson current through quantum point contacts has been
investigated in a number of papers since the pioneer work by
Beenakker and van Houten\cite{Beenakker91}. The theory describing
a single-mode quantum point contact between two s-wave
superconductors in a site representation has been developed in
\cite{LevyYeyati94,LevyYeyati95,LevyYeyati96}. In particular, in
\cite{LevyYeyati96} single-mode quantum point contact is modelled
by the only bond connecting two s-wave superconductors. At the
same time, to the best of my knowledge, the junctions between
d-wave superconductors through a few connecting bonds have not
been considered yet. Also, interference of the connecting bonds
has not been investigated by now.

The present paper is devoted to these issues. I consider two
d-wave superconductors in a mean-field site representation
connected by the only bond or by several bonds in the tunnel
limit. It is shown that for the case of several connecting bonds
their interference is very important and leads to the
non-monotonic (oscillating) behavior of the critical current in
dependence on the distance between the bonds. Furthermore, it is
found that for the $(110)$ interface-to-crystal orientation of at
least one of the superconductors the Josephson current through
each separate bond vanishes due to the symmetry and the current is
entirely determined by the interference term.

\begin{figure}[!tbh]
\centerline{\includegraphics[clip=true,width=2.1in]{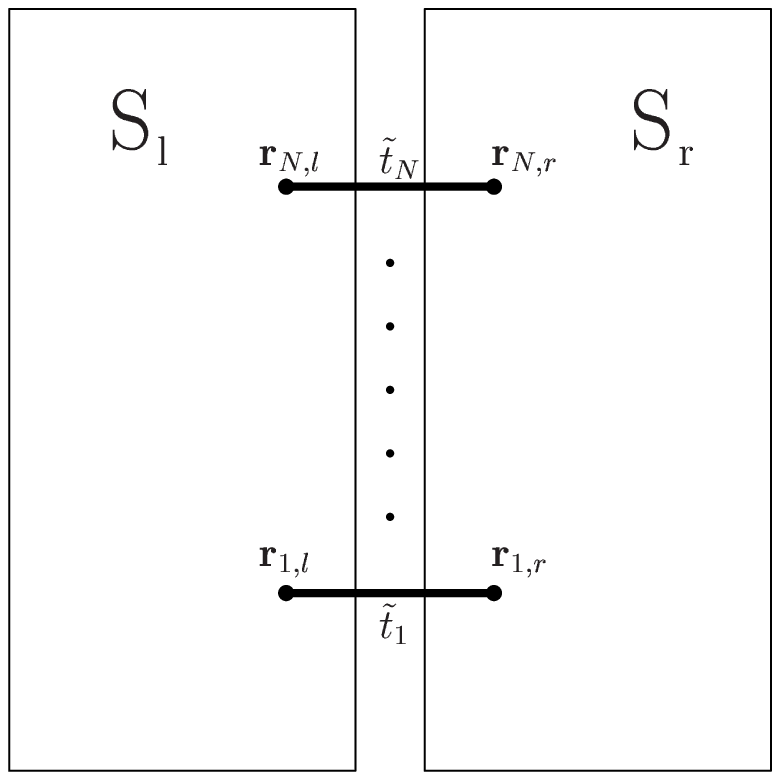}}
\caption{The scheme of the considered junction. Left and right
superconductors $S_l$ and $S_r$ are connected by the $N$ bonds
with appropriate hopping elements $\tilde t_1$,...,$\tilde t_N$}
\label{fig1}
\end{figure}

The principal scheme of the junction under consideration is shown
on the Fig.\ref{fig1}. Two superconductors are connected by the
$N$ bonds with appropriate hopping elements $\tilde
t_1$,...,$\tilde t_N$. Then the full Hamiltonian of the system
takes the form
\begin{equation} {\cal H}={\cal H}_l+{\cal H}_r+{\cal V},
\label{fham}
\end{equation}
where ${\cal H}_{l,r}$ correspond to the separate half-spaces and
$ {\cal V} $ contains coupling between them. For the each separate
half-space we can use the usual mean-field lattice Hamiltonian in
the tight-binding model
\begin{eqnarray}
{\cal H}_{l,r} =-\sum_{{\bf r},\sigma} \mu c^\dagger_{\sigma}({\bf
r}) c_{\sigma}({\bf r})- \sum_{{\bf r}\ne{\bf r'},\sigma} t_{{\bf
r},{\bf
r'}}c^\dagger_{\sigma}({\bf r}) c_{\sigma}({\bf r'})+\nonumber \\
+ \sum_{{\bf r},{\bf r'}} \{ \Delta_{{\bf r},{\bf r'}}
c^\dagger_{\uparrow}({\bf r}) c^\dagger_{\downarrow}({\bf r'}) +
h.c. \}. \label{hsham}
\end{eqnarray}
Here $\mu$ is the chemical potential; $t|_{{\bf r}\ne{\bf r}}$ are
the hopping elements. For $d$-wave superconductors pairing is
assumed to be nonzero only for the nearest neighbors $\Delta_{{\bf
r},{\bf r}\pm {\bf a}}=-\Delta_{{\bf r},{\bf r}\pm {\bf
b}}=\Delta_0$. Here ${\bf r}$ - are the site positions; ${\bf
a,~b}$ - are the basic lattice vectors ($|{\bf a}|=|{\bf b}|=1$).
The coupling ${\cal V}$ between the superconductors is modelled by
\begin{equation} {\cal V}=-\sum_{\sigma,i=1...N} \{ \tilde t_{{\bf
r_{i,l}},{\bf r_{i,r}}}c^\dagger_{\sigma}({\bf r}) c_{\sigma}({\bf
r'})+h.c.\}. \label{dist}
\end{equation}
As we are interested in the Josephson current in such a system,
the phase difference $\phi$ between the superconductors should be
taken into account. It is convenient to include it into the
coupling term ${\cal V}$:
\begin{equation}
\begin{array}{l}\tilde t_{{\bf r_{i,l}},{\bf r_{i,r}}}=\tilde t_i~e^{i\phi /2},\\
\tilde t_{{\bf r_{i,r}},{\bf r_{i,l}}}=\tilde t_i~e^{-i\phi /2}.
\end{array}\label{hopping}
\end{equation}
Then we can assume the superconducting gaps $\Delta_{l,r}$ and
hopping elements $\tilde t_i$ to be real numbers in the whole
system.

Let $\hat G({\bf r},{\bf r'})\equiv \hat G({\bf x},{\bf
x'},\omega_m)$ be the Green's function of the whole system; $\hat
G_0({\bf r},{\bf r'})$ is the Green's function of the uncoupled
superconductors; $\hat G_b({\bf r},{\bf r'})=\hat G_b({\bf r}-{\bf
r'})$ is the Green's function in the bulk of the superconductors.
Then we can represent $\hat G$ via $\hat G_0$ and $\hat T$-matrix:
\begin{equation}
\hat G({\bf r},{\bf r'})=\hat G_0({\bf r},{\bf
r'})\!+\!\!\!\!\!\!\!\sum_{\alpha,\beta=l,r \atop i,j=1...N}
\!\!\!\!\!\hat G_0({\bf r},{\bf r'_{i,\alpha}}) \hat
T_{i,j}^{\alpha,\beta} \hat G_0({\bf r_{j,\beta}},{\bf
r'}).\label{G_G0}
\end{equation}

Here all elements are $2\times2$ matrices in particle-hole space.
Then it can be derived making use of Eq.(\ref{dist}) that
$T$-matrix obeys the following equation:
\begin{equation}
\hat T_{i,j}^{\alpha,\beta}-\hat t_i^\alpha \sum_{\gamma=l,r \atop
k=1...N}\hat G_0 ({\bf r}_{i,\bar \alpha},{\bf r}_{k,\gamma}) \hat
T_{k,j}^{\gamma,\beta}=\hat t_i^\alpha \delta_{i,j} \delta_{\bar
\alpha,\beta}.\label{Tmatrix}
\end{equation}
Here $\bar l=r$, $\bar r=l$, and matrices $\hat t_i^\alpha$ are
defined by
\begin{equation}
\hat t_i^l=\left( \hat t_i^r \right)^*=\left( \begin{array}{cc}
\tilde t_i e^{i\phi /2}&0\\0& -\tilde t_i e^{-i\phi /2}
\end{array}\right).
\end{equation}
By the definition the Green's function $\hat G_0({\bf r},{\bf
r'})$ corresponds to the uncoupled superconducting half-spaces.
Then we can write
\begin{equation}
\hat G_0({\bf r}_l,{\bf r}_r)=0,~~~\hat G_0({\bf r}_r,{\bf
r}_l)=0,\label{unconnect}
\end{equation}
for any ${\bf r}_l\in S_l$ and ${\bf r}_r\in S_r$. After doing
this, we can resolve Eq.(\ref{Tmatrix}) in $(l,r)$-space:
\begin{eqnarray}
\left(\begin{array}{cc}\check T^{r,r}&\check T^{r,l}\\\check
T^{l,r}&\check T^{l,l}\end{array}\right)=~~~~~~~~
~~~~~~~~~~~~~~~~~~~~~~~~~~\nonumber \\
\left(\begin{array}{cc}\check G_0^l(1-\check G_0^r \check
G_0^l)^{-1}\check t&(1-\check G_0^l \check G_0^r)^{-1}\check t^*
\\ (1-\check G_0^r \check G_0^l)^{-1}\check t&\check
G_0^r(1-\check G_0^l \check G_0^r)^{-1}\check t^*
\end{array} \right).\label{Tresolve}
\end{eqnarray}
Here symbols with the check are $2N\times 2N$ matrices in
particle-hole and $i,j=1...N$ spaces. Namely $\check
T^{\alpha,\beta}\equiv \hat T^{\alpha,\beta}_{i,j}$, $\check
t=\hat t^l_i \delta_{i,j}$ and $\check G_0^{\alpha}\equiv \hat
t_i^{\bar \alpha} \hat G_0({\bf r}_{i,\alpha},{\bf
r}_{j,\alpha})$.

Let the $y$-axis be along the surface in the $(a,b)$-crystal plane
and the $x$-axis be the normal to surface. Then the functions
$\hat G_{0~i,j}^\alpha\equiv \hat G_0({\bf r}_{i,\alpha},{\bf
r}_{j,\alpha})$  can also be obtained from $\hat G_b$ by the $\hat
T$-matrix technique\cite{hirsh04}:

\begin{eqnarray} &&\hat{G_0}(x, x^{\prime}, k_y) = \hat{G_b}
(x - x^{\prime}, k_y) -
\\ \nonumber
&&\ \ \ \ \ \ \  \ \ \hat{G_b}(x-x_0, k_y, \omega)
\left[\hat{G_b}(0, k_y)\right]^{-1} \hat{G_b}(x_0-x^{\prime},
k_y). \label{gn}
\end{eqnarray}

Here $x_0$ is the position of isolating barrier between the
superconductors (see \cite{hirsh04} for details) and $\hat G_b(x,
k_y)$ is the Fourier transform with respect to $k_x$ of the bulk
Green's function $\hat G_b({\bf k})$ ($\hbar=1$):

\begin{eqnarray}
\hat{G}_b(n, k_y) =\frac{d}{2\pi}
\!\!\int\limits_{k_x=-\pi/d}^{\pi/d}\!\! \hat G_b (k_x, k_y) e^{i
k_x x d} dk_x\ . \label{gtilde}
\end{eqnarray}
I  only consider two possible interface-to-crystal orientations of
superconductors: $(100)$ and $(110)$. Then $d=1$ for $(100)$
orientation and $d=1/\sqrt2$ for $(110)$ orientation. Instead of
the usual square Brillouin zone $k_a = [-\pi, \pi]$, $k_b = [-\pi,
\pi]$ I now use the surface-adapted Brillouin zone given by $k_x
=[-\pi/d, \pi/d]$ and $k_y = [-\pi d, \pi d]$. Then for the
following calculations we should transform the Green's function
$\hat{G_0}(x, x^{\prime}, k_y)$ into the full coordinate
representation $\hat{G_0}({\bf r}, {\bf r}^{\prime})=\hat{G_0}(x,
x^{\prime}, y-y^{\prime})$.

The Josephson current through the barrier can be expressed by ($G$
- is the upper left part of $\hat G$, $\omega_m=(2 m+1) \pi T$)
\begin{eqnarray}
J(\phi)=-2ieT\sum_{\omega_m}\sum_{i=1...N}\left( \tilde t_{{\bf
r_{i,l}},{\bf r_{i,r}}} G({\bf r_{i,r}},{\bf r_{i,l}})-\right.\nonumber\\
\left.\tilde t_{{\bf r_{i,r}},{\bf r_{i,l}}} G({\bf r_{i,l}},{\bf
r_{i,r}})\right).~~~~~~~~~ \label{tok}
\end{eqnarray}

Taking into account Eq.(\ref{unconnect}) it can be obtained from
Eq.(\ref{G_G0}) that
\begin{equation}
\begin{array}{l}\hat G({\bf r_{i,r}},{\bf r_{i,l}})=\sum\limits_{j=1...N}
\hat G^r_{0~i,j}\hat T_{j,j}^{r,l}\hat G^l_{0~j,i}\\
\hat G({\bf r_{i,l}},{\bf r_{i,r}})=\sum\limits_{j=1...N} \hat
G^l_{0~i,j}\hat T_{j,j}^{l,r}\hat G^r_{0~j,i}.
\end{array}\label{JtrG}
\end{equation}

Then for calculating the Josephson current we  only need
$T^{l,r}_{i,i}$ and $T^{r,l}_{i,i}$ elements of $\hat T$-matrix.

Let us consider now the case of one connecting bond (i.e. $N=1$).
Then Eq.(\ref{tok}) can be easily writen explicitly:
\begin{equation}
J(\phi)=2ieT\sum_{\omega_m}\tilde t^2 \frac{F_0^r\bar F_0^l
e^{i\phi}-\bar F_0^r F_0^l e^{-i \phi}}{Z(\tilde t,\phi)},
\label{1channel}
\end{equation}
where $ F_0^{l,r}$ and $\bar F_0^{l,r}$ are the off-diagonal
elements of $\hat G_0^l$ and $\hat G_0^r$ in particle-hole space:
\begin{equation}
\hat
G_0^{\alpha}=\left(\begin{array}{cc}G_0^{\alpha}&F_0^{\alpha}\\
\bar F_0^{\alpha}& \bar G_0^{\alpha}
\end{array}\right). \label{ph-form}
\end{equation}
The denominator in Eq.(\ref{1channel}) takes the form
\begin{eqnarray}
Z(\tilde t,\phi)=1-\tilde t^2(G_0^r G_0^l+\bar G_0^r \bar
G_0^l-F_0^r\bar F_0^l e^{i\phi}-\nonumber \\\bar F_0^r F_0^l
e^{-i\phi})+ \tilde t^4(G_0^r \bar G_0^r-F_0^r\bar F_0^r)(G_0^l
\bar G_0^l-F_0^l\bar F_0^l). \label{perenorm}
\end{eqnarray}

This denominator leads to high-order powers of transparency (more
then first order in $D\sim|\tilde t|^2$) and is responsible for
the deviation of the Josephson current from the sinusoidal
behavior $J(\phi)\sim \sin(\phi)$. But for the particular problem
it is more important to consider the numerator of
Eq.(\ref{1channel}). It includes the anomalous Green's functions
of the coinciding space arguments $F_0^{\alpha}\equiv F_0({\bf
r}^{\alpha},{\bf r}^{\alpha})$ and $\bar F_0^{\alpha}$  (with
$\alpha =l,r$), for uncoupled superconductors. Here ${\bf r}^{l}$
and ${\bf r}^{r}$ - are left and right ends of the bond connecting
two superconductors. But it is easy to obtain, that for d-wave
superconductor with $(110)$ (i.e. $45^o$) smooth surface these
Green's functions are zero: $F({\bf r},{\bf r})=\bar F({\bf
r},{\bf r})=0$. This takes place due to the symmetry of sites
positions and appropriate hopping elements with respect to the
reflection $(y-y_i)\to -(y-y_i)$ near the considered site
$(x_i,y_i)$ and simultaneous changing sign of order parameter
under the reflection. These symmetry relations result in
impossibility of flowing the Josephson current through one-bond
contact  if at least one of the superconductors is  a d-wave
superconductor with $(110)$ orientation.

The vanishing of one-bond Josephson current between $(110)$
d-superconductors is a consequence of d-wave symmetry of order
parameter and has no analogue for the junctions between s-wave
superconductors, where the current is nonzero for one-bond
contact.

Any reasons which do not change the above symmetry of the system
(for example, surface pair breaking) cannot change this statement.
But if the symmetry is destroyed, the nonzero one-bond Josephson
current arises. If asymmetry is small, then current is small also.
The possible reasons, giving the nonzero Josephson current through
one-bond contact between $(110)$ d-wave superconductors are: 1)
not smooth surface of the superconductor (ends of facets or
surface roughness) at the distance of the order of $\xi_0$ from
the bond between the superconductors; 2) the impurities in the
bulk or at the surface of the superconductor, also placed not far
then $\xi_0$ from the bond; 3) nonzero $is$-component of the order
parameter or magnetic field.

\begin{figure*}[!tbh]
\begin{minipage}[b]{.5\linewidth}
   \centerline{\includegraphics[clip=true,width=3in]{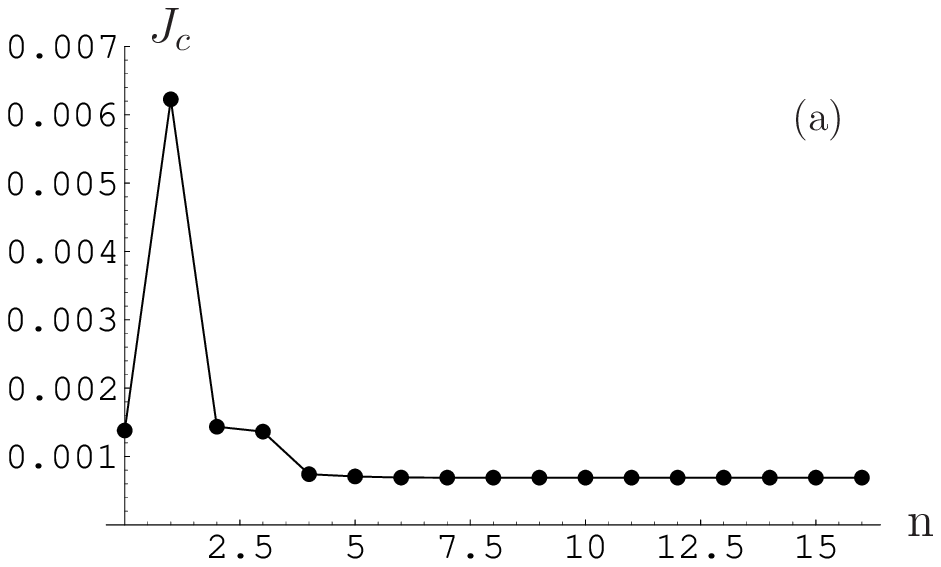}}
  \end{minipage}\hfill
  \begin{minipage}[b]{.5\linewidth}
   \centerline{\includegraphics[clip=true,width=3in]{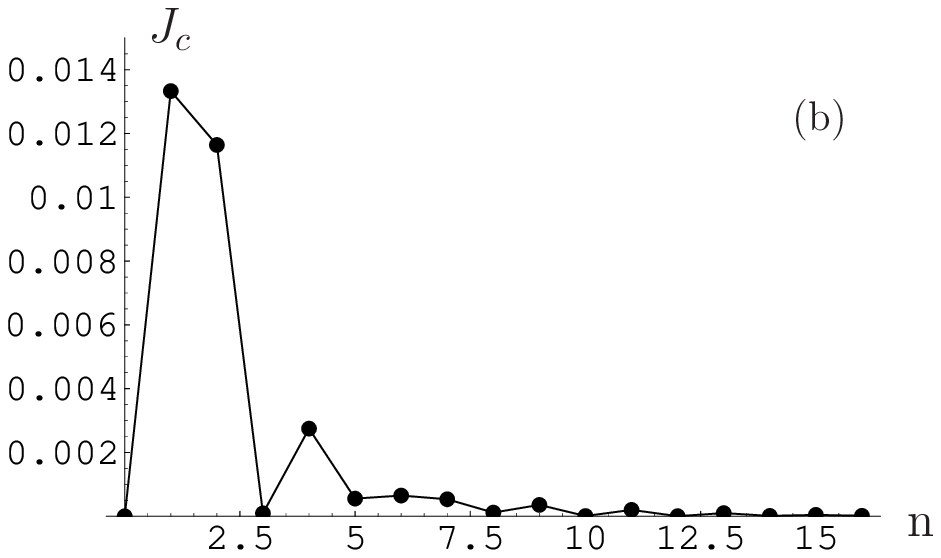}}
  \end{minipage}\hfill
\caption{The critical Josephson current in two-bond contact (in
units of $e\Delta_0(T=0)/\hbar$) plotted as a function of the
distance $L=nd_y$ between the bonds.  $T=0.6T_c$, where $T_c$ is
the superconducting critical temperature. (a) Junction between
superconductors with $(100)$ orientation (b) The same for $(110)$
orientation.} \label{fig2}
\end{figure*}

Let us consider the tunnel limit, i.e. only take into account the
first order of the barrier transparency $D\sim \tilde t^2$. This
means that values $\tilde t_i$ are sufficiently small and we can
neglect $\check G^{\alpha}_0$ in comparison with $\check 1$ in
Eq.(\ref{Tresolve}). It is worth to note that this approximation
fails at sufficiently low temperatures due to the divergence of
the Green's function $\hat G_0$ at $\omega_m\to 0$ if the
particular surface-to-crystal orientation of d-wave superconductor
leads to the formation of zero-energy surface bound states.

In the tunnel limit
\begin{equation}
\check T^{r,l}=\check t^*,~~~\check T^{l,r}=\check t.
\label{Ttunnel}
\end{equation}
Then the expression for the Josephson current takes the form:
\begin{eqnarray}
J(\phi)=2ieT\sum_{\omega_m}\sum_{i,j=1...N}\tilde t_i \tilde
t_j\left( F^r_{0~i,j}\bar F^l_{0~j,i} e^{i\phi}-\right.\nonumber \\
\left.\bar F^r_{0~j,i} F^l_{0~i,j} e^{-i
\phi}\right).~~~~~~~~~~~~~ \label{tunnTok}
\end{eqnarray}
It can be seen that Eq.(\ref{tunnTok}) contains two physically
different parts. First of them is the sum of the terms, which are
proportional to $\tilde t_i^2$. It represents the simple algebraic
sum of currents through each bond separately (compare with
Eq.(\ref{1channel})). But the other terms are proportional to
$\tilde t_i \tilde t_j$ and give the interference part of the
current. For the case of the junction with $(110)$-orientation the
interference part is the only non-vanishing term.

Now I turn to the case  $N=2$. This is the simplest case for
studying of the  effects caused by the interference  between the
connecting bonds. The results are presented for junctions between
d-wave superconductors with $(100)$ and $(110)$ orientations. In
following the superconducting order parameter is assumed to be
spatially constant. Although surface pair breaking is large for
$(110)$ surface orientation, this simplification does not change
my results qualitatively.

In  order to calculate the Josephson current Eq.(\ref{tunnTok}) we
need the Green's functions $F(x,x',y-y')$ and $\bar F(x,x',y-y')$.
They are obtained in the model of nearest neighbors with the
parameters $\Delta_0(T=0)=0.1 t$, $\mu=0.5 t$. For this set of
parameters $T_c\approx 0.173t$ and $\Delta_{max}(T=0)\approx 0.35
t$. I present the results for $T\approx 0.6 T_c$, where
$\Delta_0(0.6 T_c)\approx 0.9 \Delta_0(T=0)$. The hopping
parameters for the tunneling bonds are taken to be equal $\tilde
t_1=\tilde t_2=0.1 t$ and the bonds are placed at the distance
$L=n d_y$ from each other. Here $d_y=d^{-1}$ is the period of
lattice along the surface. For $(100)$ orientation $d_y=1$ and for
$(110)$ orientation $d_y=\sqrt 2$. Bonds $\tilde t$  connect the
last surface layers of sites of both superconductors, therefore we
should take $x=x'=x_l^{surf}$ or $x=x'=x_r^{surf}$, where
$x_{l,r}^{surf}$ are the $x$-coordinates of the positions of the
last layers in left/right superconductor. The value $y-y'$ equals
$0$ for non-interference terms and $y-y'=\pm L$ for interference
terms. Then the critical Josephson current can be written as:
\begin{equation}
J_c= 8 e T \tilde t^2\sum_{\omega_m} \left(|F_0|^2+|F_L|^2\right).
\label{2bondsTok}
\end{equation}

Here the term with $F_0\equiv F(y-y'=0)$ corresponds to the
current through each bond separately, while the term with
$F_L\equiv F(y-y'=L)$ corresponds to the interference part of the
current. The critical Josephson current described by
Eq.(\ref{2bondsTok}) is presented on Fig.\ref{fig2}. As the
current for $(110)$ orientation is entirely due to the
interference term, it  goes to zero with increasing of the
distance between the bonds. At the same time for $(100)$
orientation the current tends to the value determined by the sum
of two independent bonds. The interference part of the current
oscillates and decays with increasing $L$ due to the vanishing
$|F_L|$ at $L\to\infty$. It can be seen from Fig.\ref{fig2} that
the characteristic distance between the bonds to consider them to
be independent is $\xi_0\sim t/\Delta_0$.  It is worth to note
that the interference part of the critical current is always
positive for both orientations considered.

In conclusion, in this paper I have studied the Josephson current
through a contact with small number of connecting bonds between
two d-wave superconductors. It is obtained that the Josephson
current cannot flow through one-bond contact connecting d-wave
superconductors of $(110)$ surface-to-crystal orientation with
smooth surfaces. The expression for the Josephson current in the
junction with arbitrary number of bonds in tunnel limit is found.
It is shown that the interference between the nearest bonds is
very important. The Josephson current in two-bond junction is
calculated. It is obtained that the interference of two connecting
bonds manifests itself in non-monotonic behavior of the critical
Josephson current in dependence on the distance between the bonds
and leads to the nonzero Josephson current for $(110)$
orientation.

I thank I.V. Bobkova for many fruitful discussions and acknowledge
the support by the Dynasty Foundation and by  the grants RFBR
05-02-17175 and RFBR 05-02-17731.

% ************************** REFERENCES ********************************

\end{document}